\documentclass[onecolumn, draftcls, 11pt]{IEEEtran}
\usepackage{amsmath, amssymb, amsfonts}
\usepackage{graphicx}
\usepackage{epsfig}
\usepackage{subfigure}
\usepackage{caption}
\usepackage{cases}
\usepackage{color}
\usepackage{fancyhdr}

\begin{document}

\title{Multi-Cell Random Beamforming: Achievable Rate and Degrees of Freedom Region}
\author{Hieu Duy Nguyen, \emph{Student Member, IEEE,} Rui Zhang, \emph{Member, IEEE,} and Hon Tat Hui, \emph{Senior Member, IEEE}
\thanks{Manuscript received May 27 2012; revised December 10 2012; accepted April 27 2013. This work was presented in part at {\it IEEE International Conference on Acoustics, Speech, and Signal Processing (ICASSP)}, Kyoto, Japan, March 25-30, 2012. The associate editor coordinating the review of this paper and approving it for publication was Dr. Rong-Rong Chen.}
\thanks{Copyright (c) 2012 IEEE. Personal use of this material is permitted. However, permission to use this material for any other purposes must be obtained from the IEEE by sending a request to pubs-permissions@ieee.org.}
\thanks{The authors are with the Department of Electrical and Computer Engineering, National University of Singapore (emails: \{hieudn, elezhang, elehht\}@nus.edu.sg).}
\thanks{Digital Object Identifier XXX.}}

\maketitle \thispagestyle{empty}

\vspace{-0.50in}
\pagestyle{fancy}
\renewcommand{\headrulewidth}{0.0pt} 
\fancyhead[LO]{ACCEPTED FOR PUBLICATION IN IEEE TRANSACTIONS ON SIGNAL PROCESSING}

\begin{abstract}
\emph{Random beamforming} (RBF) is a practically favourable transmission scheme for multiuser multi-antenna downlink systems since it requires only \emph{partial} channel state information (CSI) at the transmitter. Under the conventional single-cell setup, RBF is known to achieve the optimal sum-capacity scaling law as the number of users goes to infinity, thanks to the \emph{multiuser diversity} enabled transmission scheduling that virtually eliminates the intra-cell interference. In this paper, we extend the study of RBF to a more practical multi-cell downlink system with single-antenna receivers subject to the additional inter-cell interference (ICI). First, we consider the case of \emph{finite} signal-to-noise ratio (SNR) at each receiver. We derive a closed-form expression of the achievable sum-rate with the multi-cell RBF, based upon which we show an asymptotic sum-rate scaling law as the number of users goes to infinity. Next, we consider the \emph{high-SNR} regime and for tractable analysis assume that the number of users in each cell scales in a certain order with the per-cell SNR. Under this setup, we characterize the achievable degrees of freedom (DoF) (which is defined as the sum-rate normalized by the logarithm of the SNR as SNR goes to infinity) for the single-cell case with RBF. Then we extend the analysis to the multi-cell RBF case by characterizing the \emph{DoF region}, which consists of all the achievable DoF tuples for all the cells subject to their mutual ICI. It is shown that the DoF region characterization provides useful guideline on how to design a \emph{cooperative} multi-cell RBF system to achieve optimal throughput tradeoffs among different cells. Furthermore, our results reveal that the multi-cell RBF scheme achieves the ``interference-free'' DoF region upper bound for the multi-cell system, provided that the per-cell number of users has a sufficiently large scaling order with the SNR. Our result thus confirms the optimality of multi-cell RBF in this regime even without the complete CSI at the transmitter, as compared to other full-CSI requiring transmission schemes such as \emph{interference alignment}.
\end{abstract}

\vspace{-0.20in}

\setlength{\baselineskip}{1.3\baselineskip}
\newtheorem{definition}{\underline{Definition}}[section]
\newtheorem{fact}{Fact}
\newtheorem{assumption}{\underline{Assumption}}[section]
\newtheorem{theorem}{\underline{Theorem}}[section]
\newtheorem{lemma}{\underline{Lemma}}[section]
\newtheorem{corollary}{\underline{Corollary}}[section]
\newtheorem{proposition}{\underline{Proposition}}[section]
\newtheorem{example}{\underline{Example}}[section]
\newtheorem{remark}{\underline{Remark}}[section]
\newtheorem{algorithm}{\underline{Algorithm}}[section]
\newcommand{\mv}[1]{\mbox{\boldmath{$ #1 $}}}

\section{Introduction}\label{sec:intro}

\PARstart{R}ECENT studies have shown that multiuser (MU) multiple-input multiple-output (MIMO) or multi-antenna downlink systems can achieve much higher throughput than conventional single-user (SU) MIMO counterparts by simultaneously transmitting to multiple users with different spatial signatures. The sum-capacity and the capacity region of a single-cell MU-MIMO downlink system or the so-called MIMO broadcast channel (MIMO-BC) can be achieved by the non-linear ``Dirty Paper Coding (DPC)'' scheme \cite{Costa}-\cite{Weingarten}. However, DPC requires high implementation  complexity due to the non-linear successive encoding/decoding at the transmitter/receiver, and is thus not suitable for real-time applications. Other studies have proposed to use linear precoding schemes for the MIMO-BC, e.g., the block-diagonalization scheme \cite{Spencer01}, to reduce the complexity. However, like DPC, such linear precoding schemes also rely on the assumption of perfect channel state information (CSI) at the base station (BS) transmitter, which may not be valid in practical cellular systems with a large number of users. Consequently, the study of quantized channel feedback for the MIMO-BC has become an important and active area of research (see, e.g., \cite{Jindal} and the references therein).

In a landmark work \cite{Viswanath01}, Viswanath {\it et al.} introduced a single-beam ``opportunistic beamforming (OBF)'' scheme for the MISO-BC, which exploits the multiuser diversity gain and requires only partial channel feedback to the BS. Since spatial multiplexing gain can be captured by transmitting with more than one random beams, the so-called ``random beamforming (RBF)'' scheme was also described in \cite{Viswanath01} and further investigated in \cite{Sharif01}. The achievable sum-rate with RBF in a single-cell system has been shown in \cite{Sharif01}, \cite{Sharif02}, which scales identically to that with the optimal DPC scheme assuming perfect CSI as the number of users goes to infinity, for any given user's signal-to-noise ratio (SNR). Essentially, this result implies that the intra-cell interference in a single-cell RBF system can be virtually eliminated when the number of users is sufficiently large, and an ``interference-free'' MU broadcast system is realizable.

Although substantial extensions of the single-cell RBF scheme have been pursued, there is very limited work on the performance of the RBF scheme in a more realistic multi-cell system, where the inter-cell interference (ICI) becomes a dominant factor. It is worth noting that since the universal frequency reuse is more favourable in future generation cellular systems, ICI becomes a more severe issue as compared to the traditional case with only a fractional frequency reuse. A notable work is \cite{Moon01}, in which the sum-rate scaling law for the multi-cell system with RBF has been shown to be similar to the single-cell result in \cite{Sharif01}, \cite{Sharif02} as the number of per-cell users goes to infinity, regardless of the ICI. This result, albeit appealing, does not provide any insight on how to practically design RBF in an ICI-limited multi-cell system. In this paper, we therefore aim to characterize the achievable rate for the multi-cell RBF scheme by more judiciously analyzing the impacts of ICI on the system throughput, for both the finite-SNR and high-SNR regimes.

It is worth noting that the multi-cell downlink system with ICI in general can be modelled as a Gaussian interference channel (IC). However, a complete characterization of the capacity region of the Gaussian IC, even for the two-user case, is still open \cite{Etkin}. An important recent development is the so-called ``interference alignment (IA)'' technique (see, e.g., \cite{Jafar01}-\cite{Park01} and the references therein). With the aid of IA, the maximum achievable degrees of freedom (DoF), which is defined as the sum-rate normalized by the logarithm of the SNR as the SNR goes to infinity or the so-called ``pre-log'' factor, has been obtained for various IC models to provide useful insights on designing optimal transmission schemes for interference-limited MU systems. However, IA-based DoF analysis generally relies on the perfect CSI for all the intra- and inter-cell links. Furthermore, most of the studies on IA have not yet considered the practical case of a large number of users in each cell; as a result, the role of multiuser diversity in the DoF characterization remains unaddressed. Thus, in this paper, we characterize the achievable DoFs for the multi-cell system with RBF requiring only partial CSI at the BSs, when the number of per-cell users is typically a very large number. 

Besides IA-based studies for the high-SNR regime, there is a vast body of works in the literature which investigated the multi-cell cooperative downlink precoding/beamforming at a given finite user's SNR. These results are typically categorized based on two different types of assumptions on the level of BSs' cooperation. For the case of ``fully cooperative'' multi-cell systems with global transmit message sharing across all the BSs, a virtual MIMO-BC channel is equivalently formed. Therefore, existing single-cell downlink precoding techniques can be applied (see, e.g., \cite{Zhang01}-\cite{ZhangLan} and the references therein) with a non-trivial modification to deal with the per-BS power constraints as compared to the conventional sum-power constraint for the single-cell MIMO-BC case. In contrast, if transmit messages are only locally known at each BS, coordinated precoding/beamforming can be implemented among BSs to control the ICI to their best effort \cite{Dahrouj01}-\cite{Liu01}. In \cite{Shang01}-\cite{Zhang02}, various parametrical characterizations of the Pareto boundary of the achievable rate region have been obtained for the MISO-IC with coordinated transmit beamforming and single-user detection (SUD).

Moreover, there have been recent results on the asymptotic analysis for the multi-cell MIMO downlink systems based on a large-system approach. The weighted sum-rate maximization and minimum-rate maximization problems have been considered in \cite{Huh01}, \cite{Huh02} and \cite{Zakhour01}, respectively, for various multi-cell system setups. These results considered the regime in which the number of users per cell and the number of transmit antennas per BS both go to infinity at the same time with a fixed ratio. The major goal is to apply random matrix theory to obtain ``almost closed-form'' solutions that provide deterministic approximations for the optimal precoder designs at any given finite SNR. These results in turn lead to useful insights on the system design and performance optimization. Clearly, the asymptotic regime considered in the above works differs from that in this paper, where we consider the regime in which the number of per-cell users and the per-cell SNR both go to infinity by following a given prescribed order, while the number of transmit antennas at each BS is kept as a constant. 

The main results of this paper are summarized as follows.
\begin{itemize}
\item {\bf Finite-SNR Case}: We first consider the single-cell setup and show a new closed-form expression for the achievable sum-rate with RBF, based on the known distribution of the per-beam signal-to-interference-plus-noise ratio (SINR) given in \cite{Sharif01}. We then study the multi-cell RBF and derive the SINR distribution in presence of the ICI, which is a non-trivial extension of the single-cell result in \cite{Sharif01}. Based upon this new SINR distribution, we obtain a closed-form expression of the sum-rate in the multi-cell case, and characterize the asymptotic sum-rate scaling law as the number of users per cell goes to infinity, which is shown to be identical to that for the single-cell case \cite{Sharif01} without the ICI. Notice that the same scaling law for the multi-cell RBF has been obtained in \cite{Moon01} based on an approximation of the SINR distribution, while in this paper we provide a more rigorous proof of this result using the exact SINR distribution.

\item {\bf High-SNR Case}: Although the achievable rates for the multi-cell RBF have been obtained for any given SNR with arbitrary number of
users, such results do not provide any insight to the effects of the interference on the system throughput. This motivates us to investigate the multi-cell RBF for the asymptotic high-SNR regime, under the assumption that the number of users per cell scales in a given order with the per-cell SNR (a larger order indicates a higher user density in one particular cell). Under this setup, we first consider the single-cell case and derive the maximum DoF for the achievable sum-rate with RBF, which is shown to be dependent on the user density and the number of transmit antennas, and attainable with an optimal number of random beams (data streams) employed at the transmitter. The DoF analysis thus provides a succinct description of the interplay between the multiuser diversity and spatial multiplexing gains achievable by RBF. We then seek to obtain a general characterization of the DoF region for the multi-cell RBF, which consists of all the achievable DoF tuples for all the cells subject to their mutual ICI. Different from the existing DoF region characterization based on IA \cite{Jafar01} for the case of finite number of users, in this paper we address the case with an asymptotically large number of users that scales with SNR. Our results reveal that coordination among the BSs in assigning their respective number of data beams based on different per-cell user densities is essential to achieve the optimal throughput tradeoffs among different cells. Moreover, we show that the DoF region by employing the multi-cell RBF coincides with the ``interference-free'' DoF region upper bound and thus is the exact DoF region of a multi-cell downlink system, when the user densities in different cells are sufficiently large. This result is in sharp contrast with existing studies on the achievable DoF region with the full transmitter CSI obtained by schemes such as IA \cite{Jafar01}.
\end{itemize}

The rest of the paper is organized as follows. Section \ref{sec:system model} describes the multi-cell downlink system model and the multi-cell RBF scheme. Section \ref{sec:achievable sum-rate} studies the sum-rate for both single- and multi-cell RBF systems with finite SNR. Section \ref{sec:DoF region_RBF} characterizes the achievable DoF for single-cell RBF as well as the DoF region for multi-cell RBF at the high-SNR regime. Finally, Section \ref{sec:conclusions} concludes the paper.

{\it Notations}: Scalars are denoted by lower-case letters, and vectors denoted by bold-face lower-case letters. The transpose and conjugate transpose operators are denoted as $(\cdot)^T$, and $(\cdot)^H$, respectively. $\mathbb{E}[\cdot]$ denotes the statistical expectation. $\bf{Tr}(\cdot)$ represents the trace of a matrix. The distribution of a circularly symmetric complex Gaussian (CSCG) random variable with zero mean and covariance $\sigma^2$ is denoted by $\mathcal{CN}(0,\sigma^2)$; and $\sim$ stands for ``distributed as''. $\mathbb{C}^{x \times y}$ denotes the space of $x\times y$ complex matrices.

\section{System Model}\label{sec:system model}

This paper considers a multi-cell downlink system consisting of $C$ cells, each of which has a BS with $N_T$ antennas to coordinate the transmission with $K_c$ single-antenna mobile stations (MSs), $K_c\geq 1$ and $c=1,\cdots,C$. In the $c$-th cell, the $c$-th BS transmits $M_c\leq N_T$ orthonormal beams and selects $M_c$ from $K_c$ users for transmission at each time. We assume the channels to be flat-fading and constant over each transmission period of interest. For the ease of analysis, we also assume a ``homogeneous'' channel setup, in which the received baseband signal of user $k$ in the $c$-th cell is given by
\begin{align}\label{eq:signal model}
    y_{k}^{(c)} & = \mv{h}_{k}^{(c,c)}\sum_{m=1}^{M_c} \mv{\phi}_{m}^{(c)} {\it s}_m^{(c)} \notag \\ & \qquad \qquad +\sum_{l=1,~l\neq c}^{C} \sqrt{\gamma_{l,c}} \mv{h}_{k}^{(l,c)}\sum_{m=1}^{M_l} \mv{\phi}_{m}^{(l)} {\it s}_m^{(l)} +n_{k}^{(c)},
\end{align}
where $\mv{h}_{k}^{(l,c)} \in \mathbb{C}^{1\times M_l}$ is the channel vector from the $l$-th BS to the $k$-th user of the $c$-th cell; it is assumed that all the elements of $\mv{h}_{k}^{(l,c)}$ are independent and identically distributed (i.i.d.) as $\mathcal{CN}(0,1)$; $0\leq \gamma_{l,c}<1$ stands for the distance-dependent signal attenuation from the $l$-th BS to any user in the $c$-th cell, $l\neq c$, which is less than the assumed unit direct channel gain from the $c$-th BS\footnotemark; \footnotetext{This homogeneous channel setup is required to obtain the closed-form expressions for the achievable sum-rates at finite SNRs, as will be detailed in Section III. However, the DoF region analysis for the asymptotically high-SNR regime as will be given in Section \ref{sec:DoF region_RBF} of the paper can be shown to hold even without the homogeneous channel assumption, i.e., the average signal attenuation from any BS to any user of any cell can take different values. Also note that Lemma \ref{lemma:multi-cell distributions} can be extended to the case of arbitrary signal attenuation as well.} $\mv{\phi}_{m}^{(c)} \in \mathbb{C}^{M_c\times 1}$ and $\it{s}_m^{(c)}$ are the $m$-th randomly generated beamforming vector of unit norm and transmitted data symbol from the $c$-th BS, respectively; it is assumed that each BS has the total sum power, $P_T$, i.e., ${\bf Tr}\left(\mathbb{E}[\mv{s}_{c}\mv{s}_{c}^{H}] \right)$ $\leq$ $P_T$, where $\mv{s}_{c} = [\it{s}_1^{(c)},\cdots,\it{s}_{M_c}^{(c)}]^T$; it is also assumed that the background noise is additive white Gaussian noise (AWGN), denoted by $n_{k}^{(c)}\sim\mathcal{CN}(0,\sigma^2)$, $\forall k,c$. In the $c$-th cell, the total SNR, the SNR per beam, and the interference-to-noise ratio (INR) per beam from the $l$-th cell, $l\neq c$, are denoted as $\rho=P_T/\sigma^2$, $\eta_c=P_T/(M_c\sigma^2)$, and $\mu_{l,c}=\gamma_{l,c}P_T/(M_l\sigma^2)$, respectively.
						
In this paper, we consider a multi-cell RBF scheme, in which all BSs in different cells are assumed to be able to implement the conventional single-cell RBF similarly to that given in \cite{Sharif01} at the same time, which is described as follows:
\begin{itemize}
    \renewcommand{\labelitemi}{$\bullet$}
    \item In the training phase, the $c$-th BS generates $M_c$ orthonormal beams, $\mv{\phi}_{1}^{(c)}$, $\cdots$,$\mv{\phi}_{M_c}^{(c)}$, and uses them to broadcast the training signals to all users in the $c$-th cell. The total power of each BS is assumed to be distributed equally over $M_c$ beams.

    \item Each user in the $c$-th cell measures the SINR value for each of $M_c$ beams (shown in (\ref{eq:SINR}) below), and feeds it back to the corresponding
        BS.
        \begin{align}\label{eq:SINR}
            \text{SINR}_{k,m}^{(c)}
            &= \frac{  \displaystyle\frac{P_T}{M_c}\left| \mv{h}_{k}^{(c,c)} \mv{\phi}_{m}^{(c)}\right|^2   }
          { \sigma^2 + \displaystyle\frac{P_T}{M_c}\displaystyle\sum_{i=1,i\neq m}^{M_c} \left| \mv{h}_{k}^{(c,c)} \mv{\phi}_{i}^{(c)}\right|^2
          +\displaystyle\sum_{l=1,l\neq c}^{C}\gamma_{l,c}\frac{P_T}{M_l} \displaystyle\sum_{i=1}^{M_l} \left| \mv{h}_{k}^{(l,c)} \mv{\phi}_{i}^{(l)}\right|^2
            } \notag \\
            &= \frac{  \eta_c\left| \mv{h}_{k}^{(c,c)} \mv{\phi}_{m}^{(c)}\right|^2   }
          { 1+\eta_c\displaystyle\sum_{i=1,i\neq m}^{M_c} \left| \mv{h}_{k}^{(c,c)} \mv{\phi}_{i}^{(c)}\right|^2
          +\displaystyle\sum_{l=1,l\neq c}^{C} \mu_{l,c} \displaystyle\sum_{i=1}^{M_l} \left| \mv{h}_{k}^{(l,c)} \mv{\phi}_{i}^{(l)}\right|^2
            },
        \end{align}
    where $m=1,\cdots,M_c$.

    \item The $c$-th BS schedules transmission to a set of $M_c$ users at each time by assignning its $m$-th beam to the user with the highest
    SINR, i.e.,
    \begin{align}
        k_m^{(c)}=\text{arg}\max_{k\in \{1, \cdots,K_c\}}\text{SINR}_{k,m}^{(c)}.
    \end{align}
\end{itemize}

Thus, the achievable average sum-rate in bits-per-second-per-Hz
(bps/Hz) of the $c$-th cell is given by
\begin{align}\label{eq:c-th sum-rate}
        R_{\text{RBF}}^{(c)}
        = \mathbb{E}\left[\displaystyle\sum_{m=1}^{M_c}\log_2\left(1+ \text{SINR}_{k_m^{(c)},m}^{(c)}\right)\right]
        = M_c\mathbb{E}\left[\log_2\left(1+\text{SINR}_{k_1^{(c)},1}^{(c)}\right)\right].
\end{align}

\section{Achievable rate of multi-cell Random Beamforming: finite-SNR analysis}\label{sec:achievable sum-rate}

In this section, we study the achievable sum-rate of a $C$-cell RBF system with finite SNR. We first derive a closed-form expression of the sum-rate for the single-cell case, then extend the result to the multi-cell case subject to ICI, and finally investigate the asymptotic sum-rate scaling law as the number of users per cell goes to infinity.

\subsection{Single-Cell RBF}
We first consider the single-cell case, and drop the cell index $c$ for brevity. Thus, (\ref{eq:SINR}) and (\ref{eq:c-th sum-rate}) reduce to
\begin{align}\label{eq:SINR1}
    \text{SINR}_{k,m}=\frac{\frac{P_T}{M}\left|\mv{h}_{k}\mv{\phi}_{m}\right|^2}
    {\sigma^2+\frac{P_T}{M}\sum_{i=1,i\neq m}^M\left|\mv{h}_{k}\mv{\phi}_{i}\right|^2},
    \end{align}
    \begin{align}\label{eq:sum-rate 1}
    R_{\text{RBF}} = M\mathbb{E}\left\{\log_2\left(1+\displaystyle\max_{k \in \{1,\cdots,K\}}\text{SINR}_{k,1}\right)\right].
\end{align}

The probability density function (PDF) and cumulative distribution function (CDF) of $S := \text{SINR}_{k,m}, \forall k,m$ can be expressed as \cite{Sharif01}
\begin{align}\label{eq:pdf1}
        f_S(s)=\frac{e^{-s/\eta}}{{\left(s+1\right)}^M}\left(M-1+\frac{s+1}{\eta} \right),
    \end{align}
    \begin{align}\label{eq:cdf1}
        F_S(s)=1-\frac{e^{-s/\eta}}{{\left(s+1\right)}^{M-1}},
\end{align}
where $\eta=P_T/(M\sigma^2)$ is the SNR per beam. A closed-form expression for the sum-rate $R_{\text{RBF}}$ is then given in the following lemma.

\begin{lemma}\label{lemma:bound on sum-rate single-cell}
The average sum-rate of the single-cell RBF is given by 
\begin{align}\label{eq:bound on sum-rate single-cell} R_{\text{RBF}}
= \frac{M}{\log2}\sum_{n=1}^K(-1)^n\binom{K}{n}\left[
\left(-\frac{n}{\eta}\right)^{n(M-1)}\frac{e^{n/\eta}Ei(-n/\eta)}{\left(
n(M-1)\right)!} -
\sum_{m=1}^{n(M-1)}\left(-\frac{n}{\eta}\right)^{m-1}\frac{\left(
n(M-1)-m\right)!}{\left( n(M-1)\right)!} \right],
\end{align}
where $Ei(x)=\int_{-\infty}^{x}\frac{e^t}{t}dt$ is the exponential integral function.
\end{lemma}

\begin{proof}\label{proof:bound on sum-rate single-cell}
Please refer to Appendix \ref{sec_app:bound on sum-rate
single-cell}.
\end{proof}

\begin{remark}
It is worth making a comparison between Lemma \ref{lemma:bound on sum-rate single-cell} and other existing results in the literature as follows. Note that the sum-rate expression in Lemma \ref{lemma:bound on sum-rate single-cell} is exact, while only approximations are given in \cite{Kim01} and \cite{Park02}. Moreover, (\ref{eq:bound on sum-rate single-cell}) involves the exponential integral function, which is more efficiently computable than the Gaussian hypergeometric functions given in \cite{Kim01} and \cite{Park02}. However, the sum-rate approximations in \cite{Kim01} and \cite{Park02} can directly lead to some desired asymptotic results, e.g., the sum-rate scaling law as $K\to\infty$, while (\ref{eq:bound on sum-rate single-cell}) does not. Also note that another recently obtained expression for the sum-rate of single-cell RBF can be found in \cite{Huang01}.
\end{remark}

\subsection{Multi-Cell RBF}
For the single-cell RBF case, the SINR distributions given in (\ref{eq:pdf1}) and (\ref{eq:cdf1}) were obtained in prior work \cite{Sharif01}. Now consider the multi-cell RBF case. If $\mu_{l,c}=\eta_c$, $\forall l\in\{1,\cdots,C\}\setminus\{c\}$, it is easy to see that the SINR distributions take the same forms as (\ref{eq:pdf1}) and (\ref{eq:cdf1}). However, if $\exists l\in\{1,\cdots,C\}\setminus\{c\}$ such that $\mu_{l,c}\neq\eta_c$, then the derivation of the SINR distributions in the multi-cell case becomes a new task due to the unevenly distributed ICI. To the best of the authors' knowledge, no closed-form expressions for the PDF and CDF of the SINR in this general case are available in the literature, while only approximated ones have been obtained (see, e.g., \cite{Moon01}). Therefore, in this subsection we first show a lemma on the exact SINR distributions for the multi-cell RBF, and then use them to investigate the achievable rate of this scheme.

\begin{lemma}\label{lemma:multi-cell distributions}
In the multi-cell RBF, the PDF and CDF of the SINR
$S:=\text{SINR}_{k,m}^{(c)}$, $\forall k,m$, are given by
\begin{align}\label{eq:multi-cell PDF}
            f_S^{(c)}(s)& = \frac{e^{-s/\eta_c}}{{\left(s+1\right)}^{M_c-1}
            \displaystyle\prod_{l=1,l\neq c}^C{\left(\frac{\mu_{l,c}}{\eta_c}s+1\right)^{M_l}}} \left[ \frac{1}{\eta_c} + \frac{M_c-1}{s+1} + \displaystyle\sum_{l=1,l\neq c}^{C}\frac{M_l}{s+\frac{\eta_c}{\mu_{l,c}}} \right],
        \end{align}
        \begin{align}\label{eq:multi-cell CDF}
            F_S^{(c)}(s)=1-\frac{e^{-s/\eta_c}}{{\left(s+1\right)}^{M_c-1}
            \displaystyle\prod_{l=1,l\neq c}^C{\left(\frac{\mu_{l,c}}{\eta_c}s+1\right)^{M_l}}}.
\end{align}
\end{lemma}

\begin{proof}\label{proof:multi-cell distributions}
Please refer to Appendix \ref{sec_app:multi-cell distributions}.
\end{proof}

\begin{figure}[t]
    \centering
    \epsfxsize=0.65\linewidth
  \includegraphics[width=11.5cm, height=8.5cm]{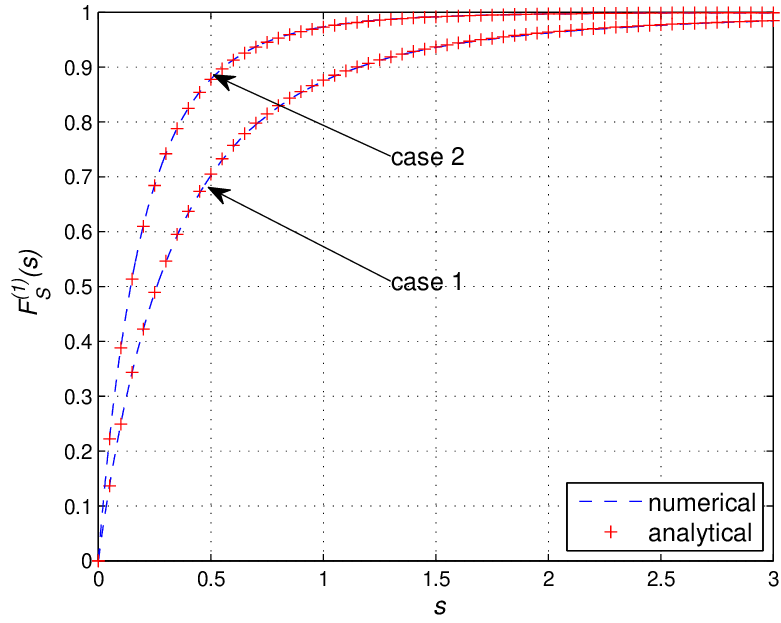}
    \caption{Comparison of the analytical and numerical CDFs of the per-cell SINR.}\label{fig:1}
    \vspace{-0.15in}
\end{figure}

It is worth noting that Lemma \ref{lemma:multi-cell distributions} nicely extends the single-cell results in (\ref{eq:pdf1}) and (\ref{eq:cdf1}), and clearly demonstrates the effect of ICI on the resulting SINR distributions. In Fig. \ref{fig:1}, we compare the analytical and numerical SINR CDFs to confirm the validity of Lemma \ref{lemma:multi-cell distributions}. We consider the SINR CDF of the first cell for two RBF systems with the following parameters: (1) $\eta_1 = 30 \text{dB}$, $M_1 = 4$, $[\mu_{2,1}, \mu_{3,1}] = [-3, 3] \text{dB}$, $[M_2, M_3] = [2, 4]$; and (2) $\eta_1 = 20 \text{dB}$, $M_1 = 6$, $[\mu_{2,1}, \mu_{3,1}, \mu_{4,1}] = [-3, 2, 3] \text{dB}$, $[M_2, M_3, M_4] = [2, 3, 4]$. It is observed that both analytical and numerical results match closely, thus justifying our derivation.

With Lemma \ref{lemma:multi-cell distributions}, Lemma \ref{lemma:bound on sum-rate single-cell} is readily generalized to the multi-cell case in the following theorem.

\begin{theorem}\label{theo:bound on sum-rate multi-cell}
Denote the total sum-rate of a $C$-cell RBF system as $\sum_{c=1}^{C}R_{\text{RBF}}^{(c)}$, where the individual sum-rate of the $c$-th cell, $R_{\text{RBF}}^{(c)}$, is given by
\begin{align}\label{eq:bound on sum-rate c-cell}
R_{\text{RBF}}^{(c)} & = \frac{M_c}{\log2}\sum_{n=1}^{K_c} (-1)^{n} \binom{K_c}{n} \prod_{l=1,l\neq c}^{C}\left(\frac{\eta_c}{\mu_{l,c}}\right)^{nM_l} \times \notag \\
& \left\{
\sum_{p=1}^{n(M_c-1)+1} \frac{A_{n,c,p}}{(p-1)!}
\left[ e^{\frac{n}{\eta_c}} \left(-\frac{n}{\eta_c}\right)^{p-1}Ei\left(-\frac{n}{\eta_c}\right) - \sum_{m=1}^{p-1}  \left(-\frac{n}{\eta_c}\right)^{m-1}(p-1-m)! \right]  \right.  \notag \\
& \left. + \sum_{l=1,l\neq c}^{C}\sum_{q=1}^{nM_l} \frac{A_{n,l,q}}{(q-1)!}
\left[ e^{\frac{n}{\mu_{l,c}}} \left(-\frac{n}{\eta_c}\right)^{q-1}Ei\left(-\frac{n}{\mu_{l,c}}\right) - \sum_{m=1}^{q-1}  \left(-\frac{n}{\eta_c}\right)^{m-1} \left(\frac{\mu_{l,c}}{\eta_c}\right)^{q-m} (q-1-m)! \right]
\right\},
\end{align}
where $A_{n,c,p}$'s and $A_{n,l,q}$'s are the coefficients from the following partial fractional decomposition:
\begin{align}\label{eq:partial fractional decomposition}
        \frac{1}{(x+1)^{n(M_c-1)+1} \prod_{l\neq c}^{C}\left( x+\frac{\eta_c}{\mu_{l,c}} \right)^{nM_l}}=
        \sum_{p=1}^{n(M_c-1)+1} \frac{A_{n,c,p}}{(x+1)^p} + \sum_{l=1,l\neq c}^{C}\sum_{q=1}^{nM_l} \frac{A_{n,l,q}}{\left( x+\frac{\eta_c}{\mu_{l,c}} \right)^q},
\end{align}
and given by \cite[2.102]{Gradshteyn}:
\begin{align}\label{eq:pfd 2}
        & A_{n,c,p} = \frac{1}{(n(M_c - 1) - p + 1)!} \frac{d^{n(M_c - 1) - p + 1}}{dx^{n(M_c - 1) - p + 1}}
        \left. \left[ \frac{1}{\prod_{l\neq c}^{C}\left( x+\frac{\eta}{\mu_l} \right)^{nM_l}} \right] \right|_{x = -1}, \\
        & A_{n,l,q} = \frac{1}{(nM_l - q)!} \frac{d^{nM_l - q}}{dx^{nM_l - q}}
        \left. \left[ \frac{1}{ (x+1)^{n(M_c-1)+1} \prod_{t\neq l, c}^{C}\left( x+\frac{\eta}{\mu_t} \right)^{nM_t}} \right] \right|_{x = -\eta_c/\mu_{l,c}}.
\end{align}
\end{theorem}

\begin{proof}\label{proof:bound on sum-rate multi-cell}
Please refer to Appendix \ref{sec_app:bound on sum-rate multi-cell}.
\end{proof}

\begin{figure}[t]
    \centering
    \epsfxsize=0.65\linewidth
  \includegraphics[width=11.5cm, height=8.5cm]{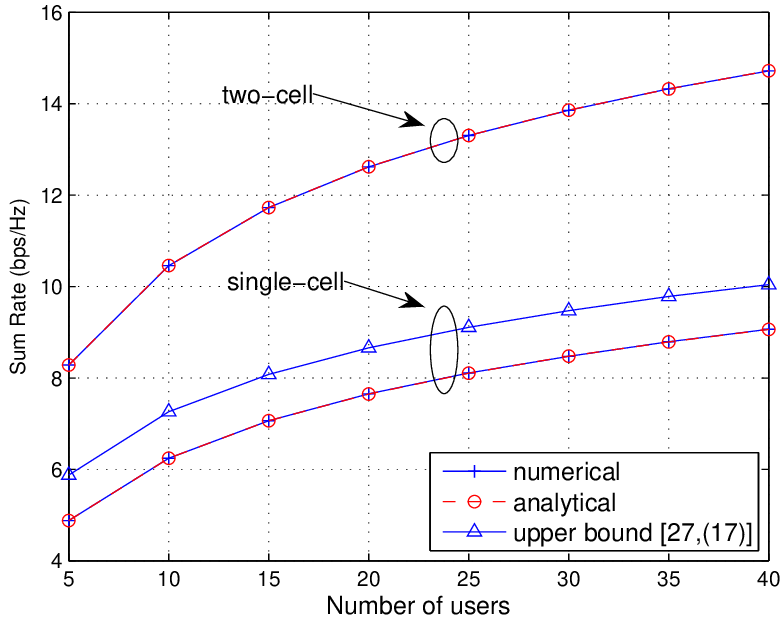}
    \caption{Comparison of the analytical and numerical results on the RBF sum-rate.}\label{fig:10}
    \vspace{-0.15in}
\end{figure}

In Fig. \ref{fig:10}, we show the analytical and numerical results on the RBF sum-rate as a function of the number of users for both single-cell and two-cell systems.We also compare the approximation obtained in \cite{Kim01}, which is only applicable to the single-cell system. In the single-cell case, $M$ = $N_T$ = 3, $\eta = 20$ dB, while in the two-cell case, $K_1$ = $K_2$, $M_1$ = $M_2$ = $N_T$ = 3, $\eta_1 = \eta_2 = 20$ dB, $\mu_{2,1} = 6$ dB, and $\mu_{1,2} = 10$ dB. It is observed that the approximation \cite[(17)]{Kim01} is only an upper bound of the achievable sum-rate. In contrast, the sum-rate expressions in (\ref{eq:bound on sum-rate single-cell}) and  (\ref{eq:bound on sum-rate c-cell}) are exact. Thus, it is feasible to use Theorem \ref{theo:bound on sum-rate multi-cell} to characterize all the sum-rate tradeoffs among different cells in a multi-cell RBF system, which leads to the achievable rate region. However, such a characterization requires intensive computations, and does not provide any useful insight. In Section \ref{sec:DoF region_RBF}, we adopt an alternative approach based on the DoF region to provide a more efficient as well as insightful tradeoff analysis for the multi-cell RBF.

\subsection{Asymptotic Sum Rate as $K_c\to\infty$}\label{subsec:conventional investigation}

It is worth noting that a conventional asymptotic investigation of RBF is to consider the case when the number of users per cell approaches infinity for a given finite SNR. Consider the single-cell RBF case with $M\leq N_T$ transmit beams and $K$ single-antenna users. The scaling law of the sum-rate is shown to be $M\log_2\log K$ as $K\to\infty$ with any fixed SNR, $\rho$, in \cite{Sharif01}, \cite{Sharif02}. An attempt to extend this result to the multi-cell RBF case has been made in \cite{Moon01} based on an approximation of the SINR's PDF (which is applicable if the SNR and INRs are all roughly equal), by showing that the same asymptotic sum-rate $M_c\log_2\log K_c$ for each individual cell holds as the single-cell case. However, we note that with the exact SINR distributions in Lemma \ref{lemma:multi-cell distributions}, a more rigorous proof can be obtained, as given in the following proposition.

\begin{proposition}\label{prop:approximated sum-rate c-cell}
For fixed $M_c$'s and $P_T$, $c=1,\cdots,C$, we have $\displaystyle\lim_{K_c\to\infty}\frac{R_{RBF}^{(c)}}{M_c\log_2(\eta_c\log K_c)}=1$.
\end{proposition}

\begin{proof}\label{proof:approximated sum-rate c-cell}
Please refer to Appendix \ref{sec_app:approximated sum-rate c-cell}.
\end{proof}

\begin{figure}[t]
    \centering
    \epsfxsize=0.65\linewidth
  \includegraphics[width=11.5cm, height=8.5cm]{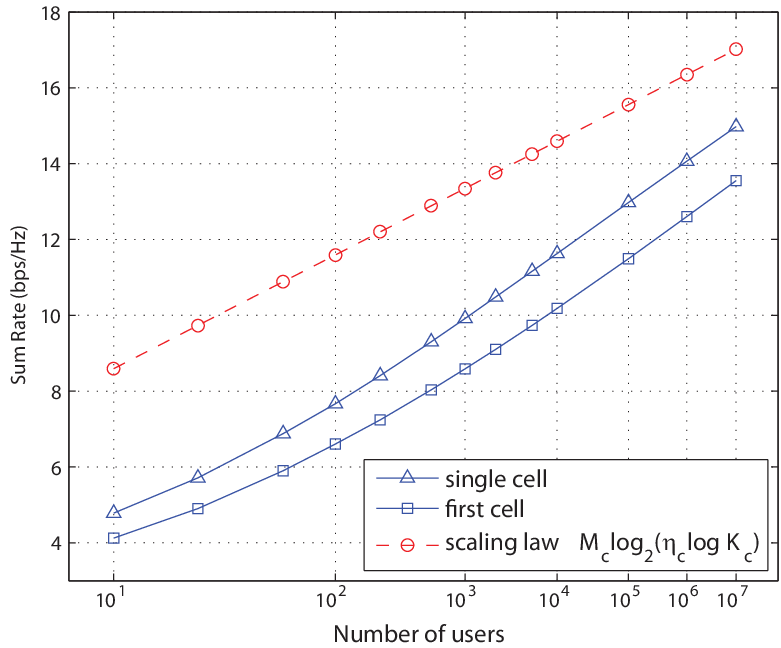}
    \caption{Comparison of the numerical sum-rate and the sum-rate scaling law for RBF.}\label{fig:10c}
    \vspace{-0.15in}
\end{figure}

In Fig. \ref{fig:10c}, we depict both the numerical and theoretical asymptotic sum-rates for a single-cell RBF system, and the first cell of a two-cell RBF system. In the single-cell case, $\eta = 5$ dB, and $M$ = $N_T$ = 3, while in the two-cell case, $M_1$ = $M_2$ = $N_T$ = 3, $\eta_1 = 5$ dB, and $\mu_{2,1} = -5$ dB. We observe that the convergence to the sum-rate scaling law $M_c\log_2(\eta_c\log K_c)$ is rather slow in both cases. For example, even with $K$ or $K_c$ to be $10^4$, the convergence is still not clearly shown. Furthermore, Proposition \ref{prop:approximated sum-rate c-cell} implies that the sum-rate scaling law $M_c\log_2(\eta_c\log K_c)$ holds for any cell regardless of the ICI as $K_c\rightarrow \infty$. As a consequence, this result implies that each BS should apply the maximum number of transmit beams, i.e., $M_c=N_T, \forall c$, to maximize the per-cell throughput. Such a conclusion may be misleading in a practical multi-cell system with non-negligible ICI. The above two main drawbacks, namely, slow convergence and misleading interpretation, have limited the usefulness of the conventional sum-rate scaling law $M_c\log_2(\eta_c\log K_c)$ for the multi-cell RBF. As will be shown in the next section, the DoF region approach is able to more precisely characterize the ICI effect on the throughput of the multi-cell RBF.

\section{Degrees of freedom region in multi-cell Random Beamforming: high-SNR analysis}\label{sec:DoF region_RBF}

In this section, we investigate the performance of the multi-cell RBF in the high-SNR regime, i.e., when the per-cell SNR $\rho\to\infty$. In particular, we consider the approach of the DoF region, which has been defined in \cite{Jafar01}, \cite{Gou01}, and is restated below for convenience.

\begin{definition}\label{def:def. DoF region}{\it (General DoF region)}
The DoF region of a $C$-cell downlink system is defined as
\begin{align}\label{eq:def. DoF region}
        \mathcal{D}=\bigg\{ (d_1,d_2,\cdots,d_C)\in\mathbb{R}_{+}^{C}:\forall
        (\omega_1,\omega_2,\cdots,\omega_C)\in\mathbb{R}_{+}^{C}  ;
        \quad \sum_{c=1}^{C}{\omega_cd_c}\leq \displaystyle\lim_{\rho\to\infty}
        \displaystyle\sup_{\mv{R}\in \mathcal{R}}
        \sum_{c=1}^{C}{\omega_c\frac{R_{sum}^{(c)}}{\log_2\rho}}
        \bigg\},
\end{align} 
where $\rho$ is the per-cell SNR; $\omega_c$, $d_c$, and $R_{sum}^{(c)}$ are the non-negative rate weight, the achievable DoF, and the sum-rate of the $c$-th cell, respectively; and the region $\mathcal{R}$ is the set of all the achievable sum-rate tuples for all the cells, denoted by $\mv{R}=(R_{sum}^{(1)}, R_{sum}^{(2)}, \cdots, R_{sum}^{(C)})$.
\end{definition}

If the multi-cell RBF is deployed, the achievable DoF region defined in (\ref{eq:def. DoF region}) is reduced to
\begin{definition}\label{def:def. DoF region_RBF}{\it (DoF region with RBF)}
The DoF region of a $C$-cell RBF system is given by
\begin{align}\label{eq:def. DoF region_RBF}
\mathcal{D}_{RBF}=\bigg\{ (d_1,d_2,\cdots,d_C)\in\mathbb{R}_{+}^{C}:\forall
        (\omega_1,\omega_2,\cdots,\omega_C)\in\mathbb{R}_{+}^{C}  ;
        \quad \sum_{c=1}^{C}{\omega_cd_c}\leq \displaystyle\lim_{\rho\to\infty}
        \displaystyle\max_{M_1, \dots, M_C \in \{ 0, \cdots, N_T\}}
        \sum_{c=1}^{C}{\omega_c\frac{R_{\text{RBF}}^{(c)}}{\log_2\rho}}
        \bigg\}.
\end{align}
\end{definition}

Certainly, $\mathcal{D}_{RBF}$ $\subseteq$ $\mathcal{D}$. Note that the DoF region is, in general, applicable for any number of users per cell, $K_c$. However, if it is assumed that all $K_c$'s are constant with $\rho\to\infty$, it can be shown that the DoF region for the multi-cell RBF given in (\ref{eq:def. DoF region_RBF}) will collapse to the null point, i.e., a zero DoF for all the cells, due to the intra-/inter-cell interference\footnotemark.\footnotetext{A rigorous proof of this claim can be deduced from Lemma \ref{lemma:theoDoF c-cell} and Theorem \ref{theorem:DoF region} later for the special case of $\alpha_c$ = 0, $\forall c$, i.e., all cells having a constant number of users.} It thus follows that for analytical tractability, the DoF region characterization for the multi-cell RBF should have $K_c$ increase in a certain order with the SNR, $\rho$. We thus make the following assumption for the rest of this paper:

\begin{assumption}\label{assump:K with rho}
The number of users in each cell scales with $\rho$ in the order of
$\rho^{\alpha_c}$, with $\alpha_c\geq 0$, denoted by
$K_c=\Theta(\rho^{\alpha_c}), c=1,\ldots, C$, i.e.,
$K_c/\rho^{\alpha_c}\to a_c$ as $\rho\to\infty$, with $a_c$ being a
positive constant independent of $\alpha_c$.
\end{assumption}

Considering the number of per-cell users to scale polynomially with the SNR is general as well as convenient. The linear scaling law, i.e., $K_c = \beta_c\rho$, with constant $\beta_c>0$, is only a special case of $K_c=\Theta(\rho^{\alpha_c})$ with $\alpha_c = 1$; if $K_c$ is a constant, then the corresponding $\alpha_c$ is zero. We can thus consider $\alpha_c$ as a measure of the user density in the  $c$-th cell given the same coverage area for all the cells, where a larger $\alpha_c$ indicates a higher number of users, $K_c$. As will be shown later in this section, Assumption \ref{assump:K with rho} enables us to obtain an efficient as well as insightful characterization of the DoF region for the multi-cell RBF. Note that the DoF region under Assumption \ref{assump:K with rho} can be considered as a generalization of the conventional DoF region analysis based on IA \cite{Jafar01} for the case of finite number of users, to the case of asymptotically large number of users that scales with the SNR. For the notational convenience, we use $\mathcal{D}(\mv{\alpha})$ and $\mathcal{D}_{RBF}(\mv{\alpha})$ to denote the achievable DoF regions $\mathcal{D}$ and $\mathcal{D}_{RBF}$, respectively, corresponding to $K_c=\Theta(\rho^{\alpha_c})$, $c$ = 1,$\cdots$,$C$, and $\mv{\alpha}$ = [$\alpha_1$, $\cdots$, $\alpha_C]^T$.

\subsection{Single-Cell Case}\label{subsec:DoF single-cell RBF}

First, we investigate the DoF for the achievable sum-rate in the single-cell RBF case without the ICI. We drop the cell index $c$ for brevity. In the single-cell case, the DoF region collapses to a line, bounded by 0 and $d_{RBF}^*(\alpha)$, where $d_{RBF}^*(\alpha)\geq 0$ denotes the maximum DoF achievable for the RBF sum-rate.

We define the achievable DoF for single-cell RBF with a given pair
of $\alpha$ and $M$ as
\begin{align}\label{eq:d(a,M)}
        d_{RBF}(\alpha,M)=\displaystyle\lim_{\rho\to\infty}\frac{R_{\text{RBF}}}{\log_2\rho}
    =\displaystyle\lim_{\eta\to\infty}\frac{R_{\text{RBF}}}{\log_2 \eta}
\end{align}
since $\eta=\rho/M$. Thus, we have $d_{RBF}^*(\alpha)$ =
$\displaystyle\max_{M\in\{ 1, \cdots, N_T\}}{d_{RBF}(\alpha,M)}$ for
a given $\alpha\geq 0$. We first characterize $d_{RBF}(\alpha,M)$ in
the following lemma.
\begin{lemma}\label{lemma:theoDoF single-cell}
    Assuming $K=\Theta(\rho^{\alpha})$, the DoF of single-cell RBF with $M\leq N_T$ orthogonal transmit beams
    is given by
    \begin{subnumcases}{d_{RBF}(\alpha,M) =}
        \frac{\alpha M}{M-1}, & $0\leq  \alpha\leq M-1$, \label{case:theoDoF single-cell a} \\
        M, & $\alpha > M-1$. \label{case:theoDoF single-cell b}
    \end{subnumcases}
\end{lemma}
\begin{proof}\label{proof:theoDoF single-cell}
Please refer to Appendix \ref{sec_app:theoDoF single-cell}.
\end{proof}

\begin{remark}
With RBF and under the assumption $K = \Theta(\rho^{\alpha})$, it is interesting to observe from Lemma \ref{lemma:theoDoF single-cell} that the achievable DoF can be a non-negative real number (as compared to the conventional integer DoF in the literature with finite $K$). Moreover, it is observed that for any given $0<\alpha<N_T-1$, assigning more transmit beams by increasing $M$ initially improves the sum-rate DoF if $M\leq \alpha+1$; however, as $M>\alpha+1$, the DoF may not necessarily increase with $M$ due to the more dominant inter-beam/intra-cell interference. Note that the term $M-1$ in the denominator of (\ref{case:theoDoF single-cell a}) is exactly the number of interfering beams to one particular data beam. Thus, Lemma \ref{lemma:theoDoF single-cell} provides a succinct description of the interplay between the available multiuser diversity (specified by $\alpha$ with a larger $\alpha$ denoting a higher user density or the number of users in a cell), the level of the intra-cell interference (specified by $M-1$), and the achievable spatial multiplexing gain or DoF, $d_{RBF}(\alpha,M)$.
\end{remark}

Next, we obtain the maximum achievable DoF for a given $\alpha$ by searching over all possible values of $M$. We note that for any $M$ $<$ $\lfloor\alpha\rfloor+1$, $d_{RBF}(\alpha,M)$ $<$ $d_{RBF}(\alpha,\lfloor\alpha\rfloor+1)$, while for any $M$ $>$ $\lfloor\alpha\rfloor+2$, $d_{RBF}(\alpha,M)$ $<$ $d_{RBF}(\alpha,\lfloor\alpha\rfloor+2)$. Thus we only need to compare $d_{RBF}(\alpha,\lfloor\alpha\rfloor+1)$ and $d_{RBF}(\alpha,\lfloor\alpha\rfloor+2)$ in searching for the optimal M. The result is shown in the following theorem.

\begin{theorem}\label{theorem:1}
For the single-cell RBF with $N_T$ transmit antennas and user density coefficient $\alpha$, the maximum achievable DoF and the corresponding optimal number of transmit beams are\footnotemark \footnotetext{The notations $\lfloor\alpha\rfloor$ and \{$\alpha$\} denote the integer and fractional parts of a real number $\alpha$, respectively.}
\begin{align}\label{eq:single dstar}
        d_{RBF}^*(\alpha) &=
            \begin{cases}
            \lfloor\alpha\rfloor+1, & \alpha\leq N_T-1, 1\geq \{\alpha\}(\lfloor\alpha\rfloor+2), \\
            \frac{\alpha(\lfloor\alpha\rfloor+2)}{\lfloor\alpha\rfloor+1},
            & \alpha\leq N_T-1, \{\alpha\}(\lfloor\alpha\rfloor+2)>1, \\
            N_T, & \alpha> N_T-1.
            \end{cases}\\{}
        \label{eq:single Mstar}
        M_{RBF}^*(\alpha) &=
            \begin{cases}
            \lfloor\alpha\rfloor+1, & \alpha\leq N_T-1, 1\geq \{\alpha\}(\lfloor\alpha\rfloor+2), \\
            \lfloor\alpha\rfloor+2, & \alpha\leq N_T-1, \{\alpha\}(\lfloor\alpha\rfloor+2)>1, \\
            N_T, & \alpha> N_T-1.
            \end{cases}
\end{align}
\end{theorem}

\begin{figure}[p]
    \centering
    \epsfxsize=0.65\linewidth
    \captionsetup{width=0.65\textwidth}
  \includegraphics[width=11.5cm, height=8.5cm]{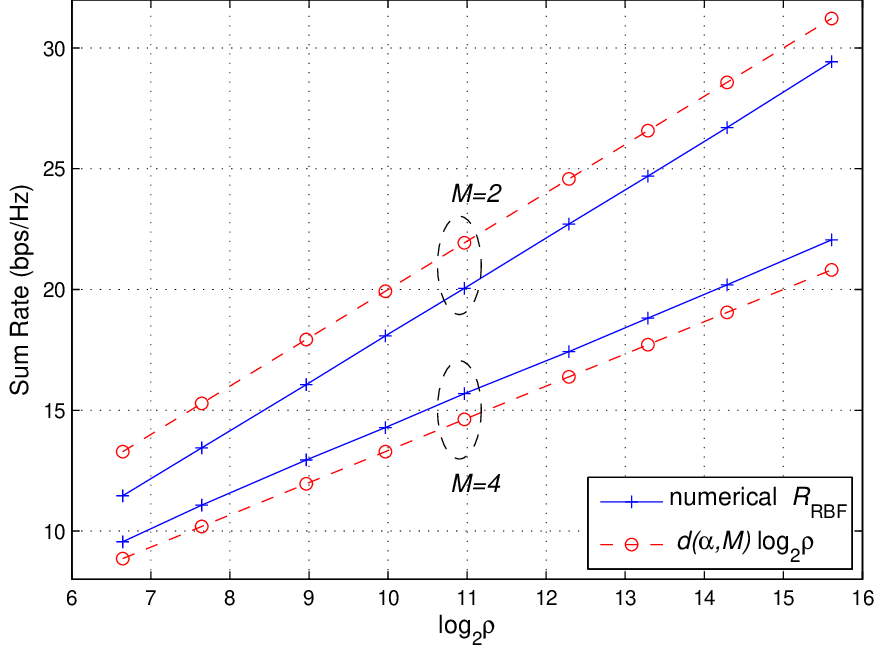}
  \vspace{-0.2in}
    \caption{Comparison of the numerical sum-rate and the scaling law $d_{RBF}(\alpha,M)\log_2\rho$, with $N_T=4$, $\alpha=1$, and $K=\lfloor \rho^{\alpha}\rfloor$.}
    \label{fig:simulation sum-rate}
    \vspace{-0.2in}
\end{figure}

In Fig. \ref{fig:simulation sum-rate}, we use simulations to confirm Lemma \ref{lemma:theoDoF single-cell}. It is observed that the newly obtained sum-rate scaling law, $R_{\text{ RBF}}=d_{RBF}(\alpha,M)\log_2\rho$, in the single-cell RBF case is very accurate, even for small values of SNR $\rho$ and number of users $K=\lfloor \rho^{\alpha}\rfloor$. Compared with Fig. \ref{fig:10c} for the conventional scaling law $R_{\rm RBF}=M\log_2(\eta\log K)$, a much quicker convergence is observed here. The DoF approach thus provides a more efficient way of characterizing the achievable sum-rate for single-cell RBF. Also observe that the sum-rate for $M=2$ is higher than that for $M=4$. This is because with $N_T=4$ and $\alpha=1$ in this example, the optimal number of beams to achieve $d_{RBF}^*(1) = 2$ is $M_{RBF}^*(1) = 2$ from (\ref{eq:single Mstar}). Since many previous studies have observed that adjusting the number of beams according to the number of users in single-cell RBF can improve the achievable sum-rate (see, e.g., \cite{Vicario01}-\cite{Kountouris01}), our study here can be considered as a theoretical explanation for such an observation.

\begin{figure}
    \centering
    \epsfxsize=0.65\linewidth
    \captionsetup{width=0.65\textwidth}
  \includegraphics[width=11.5cm, height=8.5cm]{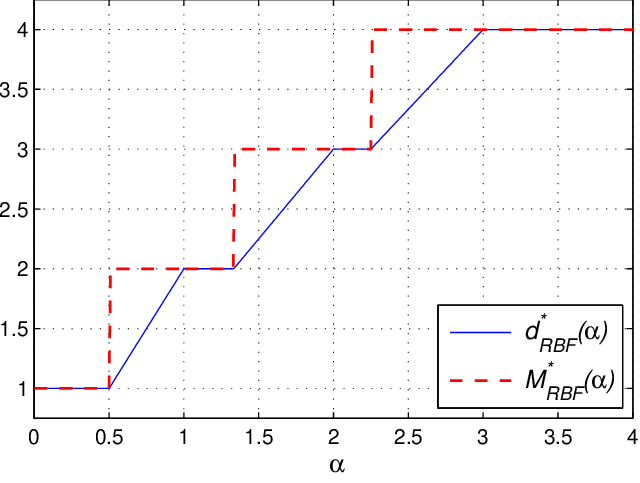}
  \vspace{-0.2in}
    \caption{The maximum DoF $d_{RBF}^*(\alpha)$ and optimal number of beams $M_{RBF}^*(\alpha)$ with $N_T=4$.}
    \label{fig:DoFsingle}
    \vspace{-0.2in}
\end{figure}

In Fig. \ref{fig:DoFsingle}, we show the maximum DoF and the corresponding optimal number of transmit beams versus the user density coefficient $\alpha$ with $N_T=4$ for single-cell RBF, according to Theorem \ref{theorem:1}. It is observed that to maximize the achievable sum-rate, we should only transmit more data beams when the number of users increases beyond a certain threshold. It is also observed that the maximum DoF $d_{RBF}^*(\alpha) = 4$ with $M=N_T = 4$ is attained when $\alpha\geq 3$ since $M_{RBF}^*(3) = 4$.

\subsection{Multi-Cell Case}\label{subsec:DoF multi-cell RBF}

In this subsection, we extend the DoF analysis for the single-cell RBF to the more general multi-cell RBF subject to the ICI. For convenience, we denote the achievable sum-rate DoF of the $c$-th cell as $d_{RBF,c}(\alpha_c,\mv{m})$ = $\lim_{\rho\to\infty}\frac{R_{\text{RBF}}^{(c)}}{\log_2 \rho}$, where $\mv{m}$ = [$M_1$,$\cdots$,$M_C]^T$ is a given set of numbers of transmit beams at different BSs. We then state the following lemma on the achieve DoF of the $c$-th cell.
\begin{lemma}\label{lemma:theoDoF c-cell}
In the multi-cell RBF, assuming
$K_c=\Theta\left(\rho^{\alpha_c}\right)$, the achievable DoF of the
$c$-th cell $d_{RBF,c}(\alpha_c,\mv{m})$, $c\in\{1,\ldots,C\}$, for
a given $\mv{m}$ is
    \begin{subnumcases}{d_{RBF,c}(\alpha_c,\mv{m}) =}
        \frac{\alpha_c M_c}{\sum_{l=1}^{C}M_l-1}, & $0\leq \alpha_c\leq \sum_{l=1}^{C}M_l-1$, \label{case:theoDoF C-cell a} \\
        M_c, & $\alpha_c > \sum_{l=1}^{C}M_l-1$. \label{case:theoDoF C-cell b}
    \end{subnumcases}\vspace{0.05in}
\end{lemma}

\begin{proof}\label{proof:theoDoF c-cell}
    The proof uses Lemma \ref{lemma:multi-cell distributions} and similar arguments in the proof of Lemma \ref{lemma:theoDoF single-cell}, and is thus omitted for brevity.
\end{proof}

\begin{remark}
Similar to Lemma \ref{lemma:theoDoF single-cell} for the single-cell case, Lemma \ref{lemma:theoDoF c-cell} reveals the relationship among the multi-user diversity, the level of the interference, and the achievable DoF for multi-cell RBF. However, as compared to the single-cell case, there are not only $M_c-1$ intra-cell interfering beams, but also $\sum_{l=1,l\neq c}^{C}M_l$ inter-cell interfering beams  for any data beam of the $c$-th cell in the multi-cell case, as observed from the denominator in (\ref{case:theoDoF C-cell a}), which results in a decrease of the achievable DoF per cell.
\end{remark}

Next, we obtain characterization of the DoF region defined in (\ref{eq:def. DoF region_RBF}) for the multi-cell RBF with any given set of per-cell user density coefficients, denoted by $\mv{\alpha}$ = [$\alpha_1$, $\cdots$, $\alpha_C]^T$ in the following theorem; for convenience, we denote $\mv{d}_{RBF}(\mv{\alpha}, \mv{m})$ = $\big[$ $d_{RBF,1}(\alpha_1, \mv{m})$ ,$\cdots$, $d_{RBF,C}(\alpha_C, \mv{m})$$\big]^T$, with $d_{RBF,c}(\alpha_c, \mv{m})$ given in Lemma \ref{lemma:theoDoF c-cell}.

\begin{theorem}\label{theorem:DoF region}
Assuming $K_c=\Theta\left(\rho^{\alpha_c}\right), c=1,\ldots,C$, the
achievable DoF region of a $C$-cell RBF system is given by
\begin{align}\label{eq:DoF region}
        \mathcal{D}_{RBF}(\mv{\alpha})=\mv{conv}& \bigg\{\mv{d}_{RBF}(\mv{\alpha}, \mv{m}),
        M_c \in \{ 0, \cdots, N_T\},~c=1,\cdots,C\bigg\},
\end{align}
where \mv{conv} denotes the convex hull operation.
\end{theorem}

Theorem \ref{theorem:DoF region} is obtained directly using Lemma \ref{lemma:theoDoF c-cell} and the definition of the DoF region, for which the proof is omitted for brevity. This theorem implies that we can obtain the DoF region of multi-cell RBF $\mathcal{D}_{RBF}(\mv{\alpha})$ by taking the convex hull over all achievable DoF points $\mv{d}_{RBF}(\mv{\alpha}, \mv{m})$ with all different values of $\mv{m}$, i.e., different BS beam number assignments.

\begin{figure}[t]
    \centering
    \epsfxsize=0.65\linewidth
  \includegraphics[width=10.5cm, height=10.5cm]{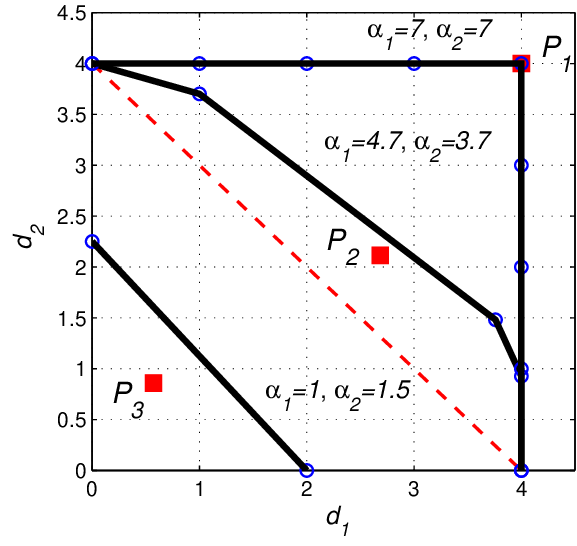}
    \caption{DoF region of two-cell RBF system with $N_T = 4$.}\label{fig:DoFregion}
    \vspace{-0.3in}
\end{figure}

In Fig. \ref{fig:DoFregion}, we depict the DoF region of a two-cell RBF system with $N_T = 4$, and for different user density coefficients $\alpha_1$ and $\alpha_2$. The vertices of these regions can be obtained by setting appropriate numbers of beams $0\leq M_1\leq 4$ and $0\leq M_2\leq 4$, while time-sharing between these vertices yields the entire boundary. To achieve the maximum sum-DoF of both cells, it is observed that a rule of thumb is to transmit more beams in the cell with a higher user density, and when $\alpha_1$ and $\alpha_2$ are both small, even turn off the BS of the cell with the smaller user density. Since the maximum sum-DoF does not consider the throughput fairness, the DoF region clearly shows all the achievable sum-rate tradeoffs among different cells, by observing its (Pareto) boundary as shown in Fig. \ref{fig:DoFregion}. It is also observed that switching the two BSs to be on/off alternately achieves the optimal DoF boundary when the numbers of users in both cells are small, but is strictly suboptimal when the user number becomes large (see the dashed line in Fig. \ref{fig:DoFregion}).

Furthermore, consider the case without any cooperation between these two BSs in assigning their numbers of transmit beams, i.e., both cells act selfishly by transmitting $M_c=N_T$ beams to aim to maximize their own DoF.The resulting DoF pairs, denoted by $\mv{d}_{RBF}([\alpha_1,\alpha_2],[4,4])$, for three sets of $\mv{\alpha}$ are shown in Fig. \ref{fig:DoFregion} as $P_1$, $P_2$, and $P_3$, respectively. It is observed that the smaller the user densities are, the further the above non-cooperative multi-cell RBF scheme deviates from the Pareto boundary. In general, the optimal DoF tradeoffs or the boundary DoF pairs are achieved when both cells cooperatively assign their numbers of transmit beams based on their respective user densities, especially when the numbers of users in both cells are not sufficiently large. Since the information needed to determine the optimal operating DoF point is only the individual cell user density coefficients, the DoF region provides a very useful method to globally optimize the coordinated multi-cell RBF scheme in practical systems.

\subsection{Optimality of Multi-Cell RBF}\label{subsec:optimality RBF}

So far, we have characterized the achievable DoF region for the multi-cell RBF scheme that requires only partial CSI at the transmitter. One important question that remains unaddressed yet is how the multi-cell RBF performs as compared to the optimal transmission scheme (e.g., IA) for the multi-cell downlink system with the full transmitter CSI, in terms of achievable DoF region. In this subsection, we attempt to partially answer this question by focusing on the regime given in Assumption \ref{assump:K with rho}, i.e., the number of users per cell scales with a polynomial order with the SNR as the SNR goes to infinity.

\subsubsection{Single-Cell Case}\label{subsubsec:optimality single-cell}

First, we consider the single-cell case to draw some useful insights. It is well known that the maximum sum-rate DoF for a single-cell MISO-BC with $N_T$ transmit antennas and $K\geq N_T$ single-antenna users with independent channels is $N_T$ \cite{Caire01}, which is achievable by the DPC scheme or even simple linear precoding schemes. However, it is not immediately clear whether such a result still holds for the case of $K=\Theta(\rho^{\alpha})\gg N_T$ with $\alpha>0$, since in this case $N_T$ may be only a lower bound on the maximum DoF. We thus have the following proposition.

\begin{proposition}\label{prop:maxDoF single-cell}
    Assuming $K=\Theta(\rho^{\alpha})$ with $\alpha > 0$, the maximum sum-rate DoF of a single-cell MISO-BC with $N_T$ transmit antennas is $d^*(\alpha) = N_T$.
\end{proposition}

\begin{proof}\label{proof:opt_single 1}
    Please refer to Appendix \ref{sec_app:maxDoF single-cell}.
\end{proof}

Proposition \ref{prop:maxDoF single-cell} confirms that the maximum DoF of the MISO-BC is still $N_T$, even with the asymptotically large number of users that scales with SNR, i.e., multiuser diversity does not yield any increment of DoF. Since from Section \ref{subsec:DoF single-cell RBF}, we know that the DoF region for the single-cell RBF scheme is a line, which is bounded by 0 and $d_{RBF}^*(\alpha)$, where $d_{RBF}^*(\alpha)$ is specified in Theorem \ref{theorem:1}. In addition, for $\alpha\geq N_T-1$, $d_{RBF}^*(\alpha) = N_T$. We thus have the following proposition.

\begin{proposition}\label{prop:optimality single-cell}
Assuming $K=\Theta(\rho^{\alpha})$, the single-cell RBF scheme is DoF-optimal if $\alpha \geq N_T-1$.
\end{proposition}

From the above proposition, it follows that the single-cell RBF achieves the maximum DoF with $M=N_T$ if the number of users is sufficiently large, thanks to the multiuser diversity effect that completely eliminates the intra-cell interference given a sufficiently large number of users.

\subsubsection{Multi-Cell Case}\label{subsubsec:optimality multi-cell}

For the convenience of analysis, we use $\mathcal{D}_{\rm UB}(\mv{\alpha})$ to denote an upper bound on the DoF region defined in (\ref{eq:def. DoF region}), for a given $\mv{\alpha}$ in the multi-cell case. Clearly, under Assumption \ref{assump:K with rho}, it follows that $\mathcal{D}_{RBF}(\mv{\alpha})$ $\subseteq$ $\mathcal{D}(\mv{\alpha})$ $\subseteq$ $\mathcal{D}_{\rm UB}(\mv{\alpha})$.

The following proposition establishes a DoF region upper bound $\mathcal{D}_{\rm UB}(\mv{\alpha})$.

\begin{proposition}\label{theorem:optimality multi-cell}
 Given $K_c = \Theta(\rho^{\alpha_c})$, $c=1,\cdots,C$, a DoF region upper bound for a $C$-cell MISO downlink system is given by
\begin{align} \label{eq:DoFregion upper bound}
    \mathcal{D}_{\rm UB}(\mv{\alpha}) = \bigg\{ (d_1,d_2,\cdots\,d_C)\in\mathbb{R}_{+}^{C}: d_c\leq N_T, c=1,\cdots,C \bigg\}.
\end{align}
\end{proposition}

The above proposition can be easily shown by noting that $d_c\leq N_T$ is a direct consequence of Proposition \ref{prop:maxDoF single-cell} for the single-cell case, which should also hold for the multi-cell case by ignoring the ICI in each of the cells, i.e., an ICI-free multi-cell downlink system is considered. Supposing $\alpha_c \geq CN_T-1$, $c$ = 1, $\cdots$, $C$, from Lemma \ref{lemma:theoDoF c-cell} and Theorem \ref{theorem:DoF region}, it easily follows that the achievable DoF region of multi-cell RBF in this case is $\mathcal{D}_{RBF}(\mv{\alpha})=\mathcal{D}_{\rm UB}(\mv{\alpha})$. This leads to $\mathcal{D}_{RBF}(\mv{\alpha})$ $\subseteq$ $\mathcal{D}(\mv{\alpha})$ $\subseteq$ $\mathcal{D}_{\rm UB}(\mv{\alpha})=\mathcal{D}_{RBF}(\mv{\alpha})$, and thus $\mathcal{D}_{RBF}(\mv{\alpha})=\mathcal{D}(\mv{\alpha})$, i.e., the multi-cell RBF achieves the exact DoF region in this regime. We thus have the following proposition.
\begin{proposition}
Given $K_c = \Theta(\rho^{\alpha_c})$, $c=1,\cdots,C$,  the
multi-cell RBF scheme achieves the DoF region of a $C$-cell MISO downlink
system, i.e.,
$\mathcal{D}_{RBF}(\mv{\alpha})=\mathcal{D}(\mv{\alpha})$, if
$\alpha_c \geq CN_T-1$, $\forall c\in \{ 1, \cdots,
C\}$.
\end{proposition}

\begin{remark} 
The above proposition implies that the multi-cell RBF is indeed DoF-optimal when the numbers of users in all cells are sufficiently large. Due to the overwhelming multiuser diversity gain, RBF compensates the lack of full CSI at transmitters without any compromise of DoF degradation. However, it is important to point out that such a result should not undermine the benefits of having the more complete CSI at transmitters in practical multi-cell systems, where more sophisticated precoding schemes than RBF such as IA-based ones \cite{Jafar01} can be applied to achieve substantial throughput gains, especially when the numbers of per-cell users are not so large. Due to the space limitation, we do not make a detailed comparison of the achievable rates between IA and RBF for the case of finite number of users in this paper, and will leave this interesting study in our further work.
\end{remark}

\section{Conclusions}\label{sec:conclusions}
In this paper, the achievable rates of the RBF scheme in a multi-cell setup subject to the ICI are thoroughly investigated. Both finite-SNR and high-SNR regimes are considered. For the finite-SNR case, we provide closed-form expressions of the achievable average sum-rates for both single- and multi-cell RBF with a finite number of users per cell. We also derive the sum-rate scaling law in the conventional asymptotic regime, i.e., when the number of users goes to infinity with a fixed SNR. Since the finite-SNR analysis has major limitations, we furthermore consider the high-SNR regime by adopting the DoF-region approach to characterize the optimal throughput tradeoffs among different cells in multi-cell RBF, assuming that the number of users per cell scales in a polynomial order with the SNR as the SNR goes to infinity. We show the closed-form expressions of the achievable DoF and the corresponding optimal number of transmit beams, both as functions of the user number scaling order or the user density, for the single-cell case. From this result, we obtain a complete characterization of the DoF region for the multi-cell RBF, in which the optimal boundary DoF point is achieved by BSs' cooperative  assignment of their numbers of transmit beams according to individual cell's user densities. Finally, if the numbers of users in all cells are sufficiently large, we show that the multi-cell RBF, albeit requiring only partial CSI at transmitters, achieve the optimal DoF region even with the full transmitter CSI. The results of this paper are useful for the optimal design of multi-cell RBF in practical cellular systems with limited channel feedback.

\appendices

\section{Proof of Lemma \ref{lemma:bound on sum-rate single-cell}}\label{sec_app:bound on sum-rate single-cell}
It is easy to see that

\vspace{-0.15in}\begin{align}
    \mathbb{E}\left[   \log_2\left(1+\displaystyle\max_{k \in \{ 1, \cdots, K\}} \text{SINR}_{k,1}\right)  \right] = \int_0^{\infty}\log_2(1+x)Kf_S(x)F_S^{K-1}(x)dx,
\end{align}
where $f_S(s)$ and $F_S(s)$ are given in (\ref{eq:pdf1}) and
(\ref{eq:cdf1}). Therefore, the RBF sum-rate can be obtained as
follows.
\begin{align}
    R_{\text{RBF}} & = M\int_0^{\infty}\log_2(1+x)Kf_S(x)F_S^{K-1}(x)dx=\frac{M}{\log2}\int_0^{\infty}\log(1+x)d\big( F_S^K(x)\big) \notag \\
    & =\frac{M}{\log2}F_S^K(x)\log(1+x)\bigg|_0^{\infty}-\frac{M}{\log2}\int_0^{\infty}\frac{1}{1+x}\left( 1 - \frac{\exp(-x/\eta)}{(1+x)^{M-1}}\right)^Kdx \notag \\
    & =\frac{M}{\log2}\lim_{x\to\infty}\big( F_S^K(x) - 1 \big)\log(1+x) + \frac{M}{\log2}\sum_{n=1}^K(-1)^n\binom{K}{n}\int_0^{\infty}\frac{\exp(-nx/\eta)dx}{(1+x)^{n(M-1)+1}} \notag \\
  & =\frac{M}{\log2}\sum_{n=1}^K(-1)^n\binom{K}{n}e^{-n/\eta}\int_1^{\infty}\frac{\exp(-ny/\eta)dy}{y^{n(M-1)+1}}.
\end{align}
Now by using \cite[2.324.2]{Gradshteyn}, (\ref{eq:bound on sum-rate
single-cell}) can be obtained. This completes the proof of Lemma
\ref{lemma:bound on sum-rate single-cell}.

\section{Proof of Lemma \ref{lemma:multi-cell distributions}}\label{sec_app:multi-cell distributions}
Denote $X=\left| \mv{h}_{k}^{(c,c)} \mv{\phi}_{m}^{(c)}\right|^2$, and
\begin{align}
    V=\displaystyle\sum_{i=1,i\neq m}^{M_c} \left| \mv{h}_{k}^{(c,c)} \mv{\phi}_{i}^{(c)}\right|^2   +\displaystyle\sum_{l=1,l\neq c}^{M_l} \frac{\mu_{l,c}}{\eta_c} \displaystyle\sum_{i=1}^{M_l} \left| \mv{h}_{k}^{(l,c)} \mv{\phi}_{i}^{(l)}\right|^2.
\end{align}

We note that the terms $\left| \mv{h}_{k}^{(l,c)}
\mv{\phi}_{i}^{(l)}\right|^2$, $\forall k, l, c, i$, are independent
chi-square random variables with two degrees of freedom, denoted by
$\chi^2(2)$. Using the characteristic function, we can express the
PDF of $V$ as follows.
\begin{align}
    f_V(v)=\frac{1}{2\pi}\displaystyle\int_{-\infty}^{\infty}{\frac{e^{-j\omega v}d\omega}
    {\left(1-j\omega\right)^{M_c-1}\displaystyle\prod_{l=1,l\neq c}^{C}
    {\left(1-j\frac{\mu_{l,c}}{\eta_c}\omega\right)^{M_l}}}},
\end{align}
where $j=\sqrt{-1}$. Since $S=\frac{X}{1/\eta_c+V}$, the PDF of $S$ is
\begin{align}
    f_{S}^{(c)}(s)=\displaystyle\int_{0}^{\infty}{f_{S|V}(s|v)f_{V}(v)dv}=A+B,
\end{align}
in which,
\begin{align}
    A=\frac{e^{-s/\eta_c}}{2\pi\eta_c}\displaystyle\int_{-\infty}^{\infty}{\frac{d\omega}
    {(s+j\omega)\left(1-j\omega\right)^{M_c-1}\displaystyle\prod_{l=1,l\neq c}^{C}
    {\left(1-j\frac{\mu_{l,c}}{\eta_c}\omega\right)^{M_l}}}},
\end{align}
\begin{align}
    B=\frac{e^{-s/\eta_c}}{2\pi}\displaystyle\int_{-\infty}^{\infty}{\frac{d\omega}
    {(s+j\omega)^2\left(1-j\omega\right)^{M_c-1}\displaystyle\prod_{l=1,l\neq c}^{C}
    {\left(1-j\frac{\mu_{l,c}}{\eta_c}\omega\right)^{M_l}}}}.
\end{align}

We first use the following partial fraction expansion to decompose $A$,
\begin{align}\label{eq:decomposition}
	\frac{1}{(s+j\omega)\left(1-j\omega\right)^{M_c-1}\displaystyle\prod_{l=1,l\neq c}^{C} {\left(1-j\frac{\mu_{l,c}}{\eta_c}\omega\right)^{M_l}}} =  \frac{A_0}{s+j\omega}
	    + \frac{A_c}{1-j\omega} + \displaystyle\sum_{l=1,l\neq c}^{C}\frac{A_l}{1-j\frac{\mu_{l,c}}{\eta_c}\omega}
	    + A',
\end{align}
where $A'$ is the sum of all terms with order greater than 1.

We then apply the {\it Cauchy integral formula} \cite{Brown01} to obtain
\begin{align}\label{eq:proof02_01}
    A &= \frac{e^{-s/\eta_c}}{2\pi\eta_c} \int_{-\infty}^{\infty} \left[  \frac{A_0}{s+j\omega}
    + \frac{A_c}{1-j\omega} + \displaystyle\sum_{l=1,l\neq c}^{C}\frac{A_l}{1-j\frac{\mu_{l,c}}{\eta_c}\omega}
    + A' \right]d\omega \notag \\
    &= \frac{e^{-s/\eta_c}}{2\eta_c} \left( A_0 + A_c + \displaystyle\sum_{l=1,l\neq c}^{C}\frac{\eta_c}{\mu_{l,c}}A_l \right).
\end{align}

Now from (\ref{eq:decomposition}) we note that 
\begin{align}\label{eq:proof02_02}
    1 &= A_0(1-j\omega)^{M_c-1}\displaystyle\prod_{l=1,l\neq c}^{C}
    \left(1-j\frac{\mu_{l,c}}{\eta_c}\omega\right)^{M_l} + A_c(s+j\omega)(1-j\omega)^{(M_c-2)^+}\displaystyle\prod_{l=1,l\neq c}^{C}\left(1-j\frac{\mu_{l,c}}{\eta_c}\omega\right)^{M_l} \notag \\
    &+ (s+j\omega)(1-j\omega)^{M_c-1}\displaystyle\sum_{l=1,l\neq c}^{C}\left[  A_l\left(  1-j\frac{\mu_{l,c}}{\eta_c}\omega  \right)^{M_l-1}\displaystyle\prod_{n=1,n\neq l, c}^{C}{\left(1-j\frac{\mu_{n,c}}{\eta_c}\omega\right)^{M_n}} \right] \notag \\
    &+ A'(s+j\omega)\left(1-j\omega\right)^{M_c-1}\displaystyle\prod_{l=1,l\neq c}^{C} {\left(1-j\frac{\mu_{l,c}}{\eta_c}\omega\right)^{M_l}}.
\end{align}
where the notation $(x)^+$ represents $\max(x,0)$. Substituting
$\omega=-s/j$, we have
\begin{align}
A_0=\frac{1}{(1+s)^{M_c-1}\prod_{l=1,l\neq c}^{C}{
(1+\frac{\mu_{l,c}}{\eta_c}s)^{M_l}  }}.
\end{align}
Also, by observing the coefficient of
$\omega^{(\sum_{l=1}^{C}{M_l}-1)}$, (\ref{eq:proof02_02}) leads to
$A_0=A_c + \sum_{l=1,l\neq c}^{C}\frac{\eta_c}{\mu_{l,c}}A_l$.
Therefore, from (\ref{eq:proof02_01}), we conclude
\begin{align}\label{eq:proof02_03}
    A=\frac{e^{-s/\eta_c}}{\eta_c}A_0=\frac{e^{-s/\eta_c}}{\eta_c(1+s)^{M_c-1}\prod_{l=1,l\neq c}^{C}{  (1+\frac{\mu_{l,c}}{\eta_c}s)^{M_l}  }}.
\end{align}

By differentiating $A$ in (\ref{eq:proof02_01}) with respect to $s$,
we obtain $\frac{dA}{ds}=-\frac{A}{\eta_c}-\frac{B}{\eta_c}$, i.e.,
$f_S^{(c)}(s) = A + B = -\eta_c\frac{dA}{ds}$. Combining this result
and (\ref{eq:proof02_03}), (\ref{eq:multi-cell PDF}) and
(\ref{eq:multi-cell CDF}) are obtained. This completes the proof of
Lemma \ref{lemma:multi-cell distributions}.

\section{Proof of Theorem \ref{theo:bound on sum-rate multi-cell}}\label{sec_app:bound on sum-rate multi-cell}
Using the similar derivation as in Appendix \ref{sec_app:bound on
sum-rate single-cell}, we obtain
\begin{align}
    R_{\text{RBF}}^{(c)} & = \frac{M_c}{\log2}\sum_{n=1}^{K_c}(-1)^n\binom{K_c}{n}\int_0^{\infty}\frac{\exp(-nx/\eta)dx}{(1+x)^{n(M_c-1)+1}\prod_{l=1,l\neq c}^C{\left(\frac{\mu_{l,c}}{\eta_c}x+1\right)^{nM_l}}} \notag \\
& = \frac{M_c}{\log2}\sum_{n=1}^{K_c}(-1)^n\binom{K_c}{n}\prod_{l=1,l\neq c}^{C}\left(\frac{\eta_c}{\mu_l}\right)^{nM_l}\int_0^{\infty}\frac{\exp(-nx/\eta)dx}{(x+1)^{n(M_c-1)+1}\prod_{l=1,l\neq c}^C{\left(x + \frac{\eta_c}{\mu_{l,c}}\right)^{nM_l}}}.
\end{align}
By applying the partial fractional decomposition given in
(\ref{eq:partial fractional decomposition}) and using
\cite[2.234.2]{Gradshteyn} for each term therein, we arrive at
(\ref{eq:bound on sum-rate c-cell}). This completes the proof of
Theorem \ref{theo:bound on sum-rate multi-cell}.

\section{Proof of Proposition \ref{prop:approximated sum-rate c-cell}}\label{sec_app:approximated sum-rate c-cell}

Due to the similarity between (\ref{eq:cdf1}) and
(\ref{eq:multi-cell CDF}), the original approach in \cite[Theorem
1]{Sharif01} can be applied to prove this proposition with minor
modifications. For the completeness, only a sketch proof is
presented here. 

To show that $R_{\text{RBF}}^{(c)} \xrightarrow{K_c\to\infty}
M_c\log_2\log K_c$, we first note that $f_S^{(c)}(s)$ and
$F_S^{(c)}(s)$ satisfies the {\it von Mises condition} for the
Gumbel-type limiting distributions (see, e.g., \cite[Theorem
10.5.2.c]{David01}). Therefore, as $K_c\to\infty$, there exist
constants $a_{K_c}$ and $b_{K_c}$ such that
$\big[F_S^{(c)}(a_{K_c}x+b_{K_c})\big]^{K_c}\to \exp\left(-e^{-x}
\right)$. The value of $b_{K_c}$ can be found to be\footnotemark
\begin{align}
    b_{K_c} = \eta_c\log K_c - \eta_c\left(\sum_{l=1}^{C}M_l - 1 \right)\log\log K_c + O(\log\log\log K_c).
\end{align}
\footnotetext{Let $f(K_c)$ be a function of $K_c$. $f(K_c) =
O(\log\log\log K_c)$ means that $f(K_c)/\log\log\log K_c < \infty$
as $K_c\to\infty$.} Furthermore, the growth function, defined as
$g_S^{(c)}(s) = \left( 1 - F_S^{(c)}(s)\right)/f_S^{(c)}(s)$ for
$s\geq 0$, is given by
\begin{align}
    g_S^{(c)}(s) = \frac{1}{\frac{1}{\eta_c} + \frac{M_c-1}{s+1} + \sum_{l=1,l\neq c}^C\frac{M_l}{s+\frac{\eta_c}{\mu_{l,c}}}}.
\end{align}

It is easy to verify the followings:
\begin{itemize}
    \item $\lim_{s\to\infty}g_S^{(c)}(s) = \eta_c > 0$,
    \item $b_{K_c} = O(\log K_c)$ as $K_c\to\infty$, and
    \item The derivative of $g_S^{(c)}(s)$ satisfies
\begin{align}
            \frac{d^n g_S^{(c)}(s)}{ds^n}\bigg|_{s = b_{K_c}} = O\left(\frac{1}{b_{K_c}^{n+1}}\right).
\end{align}
\end{itemize}

Hence, by applying \cite[Corollary A.1]{Sharif01}, we have
\begin{align}
    & \text{Pr}\left\{ \eta_c\log K_c - \eta_c \left(\sum_{l=1}^{C} M_l\right)\log\log K_c + O(\log\log\log K_c) \leq \max_{k\in\{1,\cdots,K_c\}}\text{SINR}_{k,m}^{(c)} \right.\notag \\
    & \qquad \qquad \left. \leq  \eta_c\log K_c - \eta_c\left(\sum_{l=1}^{C} M_l - 2\right)\log\log K_c + O(\log\log\log K_c)  \right\} \geq 1 - O\bigg( \frac{1}{\log K_c}\bigg).
\end{align}

This completes the proof of Proposition
\ref{prop:approximated sum-rate c-cell}.

\section{Proof of Lemma \ref{lemma:theoDoF single-cell}}\label{sec_app:theoDoF single-cell}

For convenience, we denote the following auxiliary random variable
$R_{k,m}=\log_2\left(1+\text{SINR}_{k,m}\right)$. From
(\ref{eq:cdf1}), the CDF of $R_{k,m}$ is obtained as
\begin{align}
        F_R(r)=1-\frac{e^{-\left(2^r-1\right)/\eta}}{2^{r(M-1)}}. \notag
\end{align}

To prove Lemma \ref{lemma:theoDoF single-cell}, we first show that
\begin{align}   \label{eq:proof01_1a}
    \text{Pr}\left\{\frac{\alpha}{M-1}\log_2\eta +\log_2\log\eta\geq
    \max_{k \in \{ 1, \cdots, K\}}R_{k,1} \right.
    \left. \geq \frac{\alpha}{M-1}\log_2\eta -\log_2\log\eta \right\}
    \xrightarrow{\eta\to\infty}1,~ \text{if}~ 0< \alpha\leq M-1,
\\ \label{eq:proof01_1b}
    \text{Pr}\bigg\{ \log_2\eta +\log_2\log\eta +\log_2\alpha\geq
    \max_{k \in \{ 1, \cdots, K\}}R_{k,1}
    \geq\log_2\eta +\log_2\log\eta +\log_2\beta \bigg\}
    \xrightarrow{\eta\to\infty}1,~ \text{if}~ \alpha> M-1,
\end{align}
in which the constant $\beta$ is define as
$\beta=\frac{\alpha-M+1}{2}$; hence $\alpha>\beta>0$ when
$\alpha>M-1$. Considering (\ref{eq:proof01_1a}), the upper-bound 
probability can be expressed as
\begin{align}
    \text{Pr}\left\{\frac{\alpha}{M-1}\log_2\eta +\log_2\log\eta\geq \max_{k \in \{ 1, \cdots, K\}}R_{k,1} \right\}
    &=\bigg[ F_R\left( \frac{\alpha}{M-1}\log_2\eta +\log_2\log\eta \right) \bigg]^K \notag \\
    &=\left(  1 - \frac{ \exp{\left( -\eta^{\frac{\alpha}{M-1}-1}\log\eta \right)} \exp{(\eta^{-1})} }{\eta^{\alpha}(\log\eta)^{M-1}} \right)^K.
\end{align}

Using the asymptotic relation $\log(1-x)=-x+O(x^2)$ when $x$ is
small, we get
\begin{align}\label{eq:proof01_1c}
   & K\log\left(  1 - \frac{ \exp{ \left(-\eta^{\frac{\alpha}{M-1}-1}\log\eta \right)} \exp{(\eta^{-1})} }{\eta^{\alpha}(\log\eta)^{M-1}} \right)
    = -\frac{K}{\eta^{\alpha}(\log\eta)^{M-1}} \exp{ \left(-\eta^{\frac{\alpha}{M-1}-1}\log\eta \right)} \exp{(\eta^{-1})} \notag \\
   ~~~~~~~&  + O\left( \frac{K}{\eta^{2\alpha}(\log\eta)^{2(M-1)}} \exp{ \left(-2\eta^{\frac{\alpha}{M-1}-1}\log\eta \right)} \exp{(2\eta^{-1})} \right) \xrightarrow{\eta\to\infty}0,
\end{align}
in which we have used the assumptions $K=\Theta(\eta^{\alpha})$, and
$0<\alpha\leq M-1$. As a consequence, the upper-bound probability
converges to 1 when $\eta\to\infty$. To show the convergence of the
lower-bound probability in (\ref{eq:proof01_1a}), we can utilize the
same technique described above by showing
\begin{align}\label{eq:proof01_1d}
    \text{Pr}\left\{\frac{\alpha}{M-1}\log_2\eta -\log_2\log\eta\geq \max_{k \in \{ 1, \cdots, K\}}R_{k,1} \right\}
    &=\bigg[ F_R\left( \frac{\alpha}{M-1}\log_2\eta -\log_2\log\eta \right) \bigg]^K \notag \\
    &=\left(  1 - \frac{ \exp{\left( -\frac{1}{\log\eta}\eta^{\frac{\alpha}{M-1}-1} \right) } \exp{(\eta^{-1})} (\log\eta)^{M-1} }{\eta^{\alpha}} \right)^K.
\end{align}

Note that
\begin{align}\label{eq:proof01_1e}
    & K\log\left(  1 - \frac{ \exp{\left( -\frac{1}{\log\eta}\eta^{\frac{\alpha}{M-1}-1} \right)} \exp{(\eta^{-1})} (\log\eta)^{M-1} }{\eta^{\alpha}} \right)    \notag \\
    & = -\frac{K}{\eta^{\alpha}}  (\log\eta)^{M-1}  \exp{\left( -\frac{1}{\log\eta}\eta^{\frac{\alpha}{M-1}-1} \right) }  \exp{(\eta^{-1})} \notag \\
    & \qquad \qquad
    + O\left( \frac{K}{\eta^{2\alpha}}  (\log\eta)^{2(M-1)}  \exp{\left( -\frac{2}{\log\eta}\eta^{\frac{\alpha}{M-1}-1} \right) }  \exp{(2\eta^{-1})} \right) \xrightarrow{\eta\to\infty}-\infty,
\end{align}
since, when $\eta\to\infty$, the first term in (\ref{eq:proof01_1e})
goes to $-\infty$, while the second term goes to 0.
(\ref{eq:proof01_1d}) thus converges to 0 and the lower-bound
probability is confirmed. The proof of (\ref{eq:proof01_1b}) follows
similar arguments as the above, and is omitted for brevity. This
completes the proof of Lemma \ref{lemma:theoDoF single-cell}.

\section{Proof of Proposition \ref{prop:maxDoF single-cell}}\label{sec_app:maxDoF single-cell}
According to the main text, it is sufficient to show the DoF upper
bound to be $N_T$ as follows. Note that in a single-cell MISO-BC,
the DPC yields the optimal sum-rate, denoted by $R_{\text{DPC}}$.
Therefore, $d^*(\alpha) =
\displaystyle\lim_{\rho\to\infty}\frac{R_{\text{DPC}}}{\log_2\rho}$.
From \cite[Theorem 1]{Jindal01}, we have
\begin{align}
        R_{\text{DPC}} \leq N_T\mathbb{E}\left[ \log_2\left[ 1 + \eta\max_{k \in \{ 1, \cdots, K\}}||\mv{h}_k||^2 \right] \right].
    \end{align}
Note that $\eta = P_T/(N_T\sigma^2)$ is the SNR per beam, and
$||\mv{h}_k||^2$'s are i.i.d. chi-square random variables with
$2N_T$ degrees of freedom, denoted by $\chi^2(2N_T)$. Thus, if we
denote $R_k=\log_2(1+\eta||\mv{h}_k||^2)$, the CDF of $R = R_k$ is
$F_R(r)=\frac{1}{\Gamma(N_T)}\gamma\left( N_T, \frac{2^r-1}{\eta}
\right)$, where $\Gamma(\cdot)$ and $\gamma(\cdot,\cdot)$ are the
gamma and the incomplete gamma function, respectively. The same
reasoning as in the proof of Lemma \ref{lemma:theoDoF single-cell}
can be reused here to show that
\begin{align}
        \text{Pr}\left\{ \log_2\eta + \log_2\log\eta + \log_2(\alpha+1)\geq \max_{k \in \{ 1, \cdots, K\}}R_k\right\} \xrightarrow{\eta\to\infty}1.
    \end{align}
This completes the proof of Proposition \ref{prop:maxDoF
single-cell}.



\begin{biography}[{\includegraphics[width=1in,height=1.25in,clip,keepaspectratio]{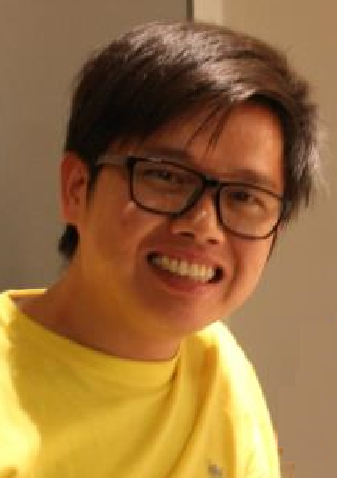}}]
{\bf Hieu Duy Nguyen} (S'10) received the B.S. (First-Class Hons.) degree in electrical engineering from the Vietnam National University, Hanoi, Vietnam, in 2009. In January 2010, he joined the Department of Electrical and Computer Engineering, National University of Singapore as a Research Scholar, where he is currently working towards his Ph.D. degree. His current research interests include wireless communications and information theory, focusing on multiuser MIMO systems, cooperative communications, and interference channels.
\end{biography}

\begin{biography}[{\includegraphics[width=1in,height=1.25in,clip,keepaspectratio]{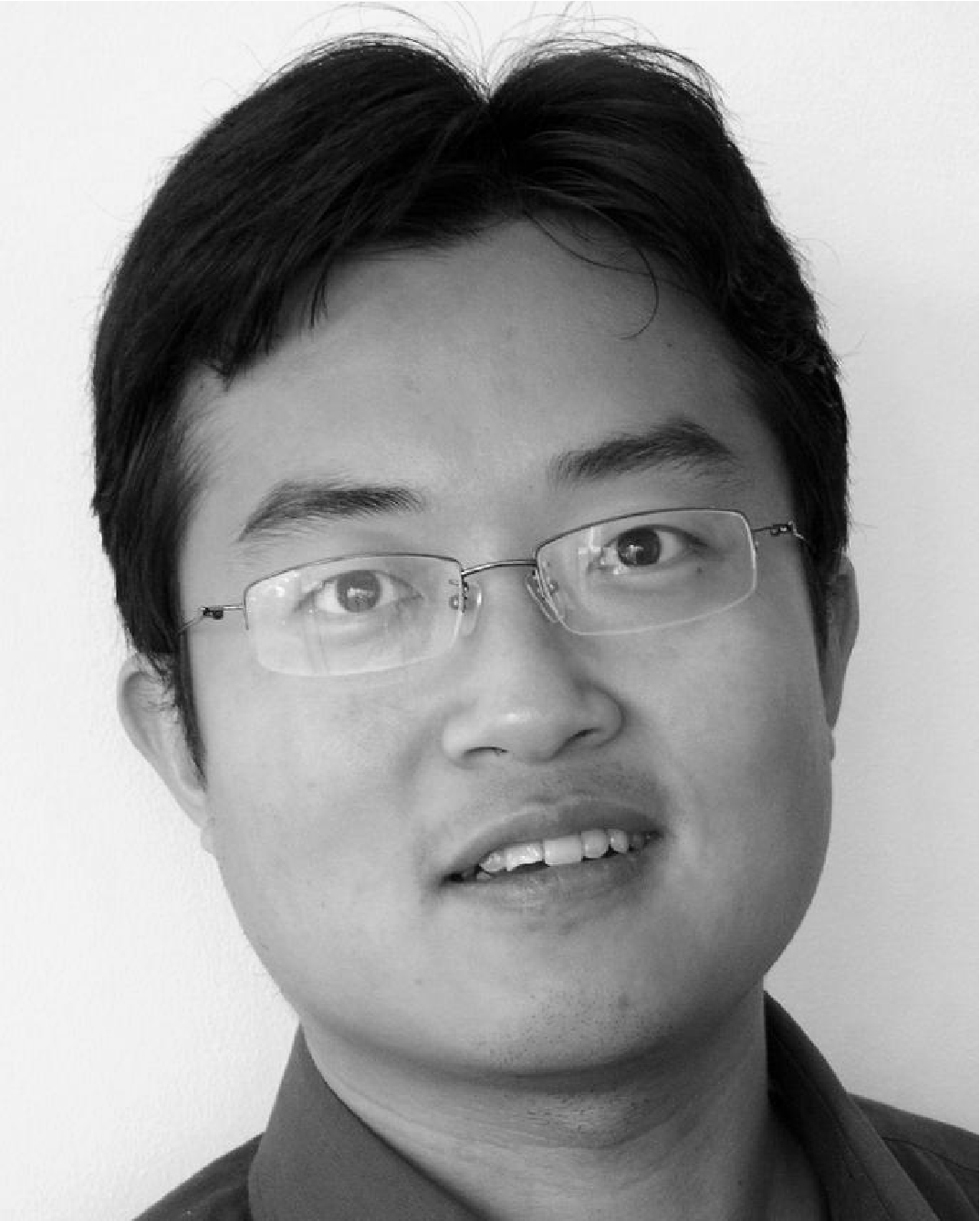}}]
{\bf Rui Zhang}(S'00--M'07) received the B.Eng. (First-Class Hons.) and M.Eng. degrees from the National University of Singapore in 2000 and 2001, respectively, and the Ph.D. degree from the Stanford University, Stanford, CA USA, in 2007, all in electrical engineering. Since 2007, he has worked with the Institute for Infocomm Research, A-STAR, Singapore, where he is now a Senior Research Scientist. Since 2010, he has joined the Department of Electrical and Computer Engineering of the National University of Singapore as an Assistant Professor. His current research interests include multiuser MIMO, cognitive radio, cooperative communication, energy efficient and energy harvesting wireless communication, wireless information and power transfer, smart grid, and optimization theory. He has authored/coauthored over 140 internationally refereed journal and conference papers. He was the co-recipient of the Best Paper Award from the IEEE PIMRC in 2005. He was the recipient of the 6th IEEE ComSoc Asia-Pacific Best Young Researcher Award in 2010, and the Young Investigator Award of the National University of Singapore in 2011. He is now an elected member of IEEE Signal Processing Society SPCOM and SAM Technical Committees, and an editor for the IEEE Transactions on Wireless Communications.
\end{biography}

\begin{biography}[{\includegraphics[width=1in,height=1.25in,clip,keepaspectratio]{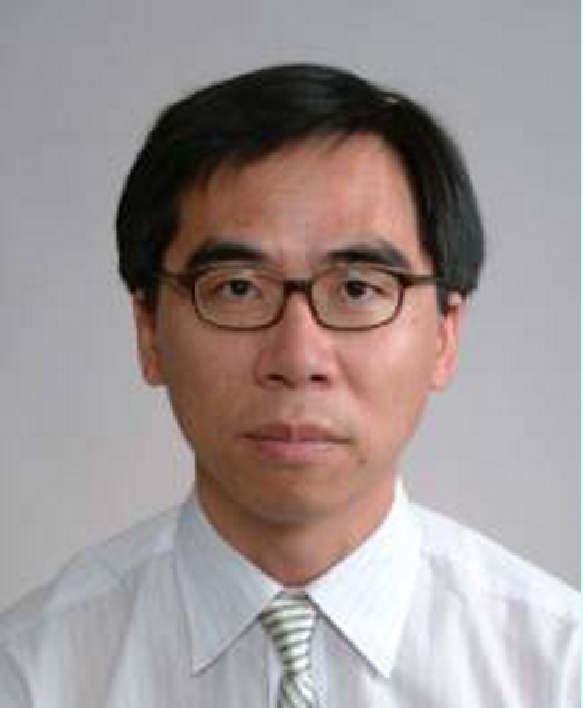}}]
{Hon Tat Hui} (SM'04) received the B.Eng. (Hons.) degree with first class honors from the City University of Hong Kong, Kowloon, Hong Kong, in 1994, and the Ph.D. degree from The City University of Hong Kong, in 1998. From 1998 to 2001, he was a Research Fellow in the City University of Hong Kong. From 2001 to 2004, he was an Assistant Professor in the Nanyang Technological University in Singapore. From 2004 to 2007, he was a Lecturer in the School of Information Technology and Electrical Engineering, University of Queensland. He is currently an Assistant Professor with the Department of Electrical and Computer Engineering, National University of Singapore. He has published more than 70 papers in internationally refereed journals. His research interest is in the use of antennas for biomedical applications. Dr. Hui has been a Project Reviewer of various industrial and government organizations. He was the exceptional performance reviewer for the IEEE Antennas and Propagation Society in 2008 and 2009. He served as an editorial board member in various international journals. He has helped organize many local and international conferences.
\end{biography}


\end{document}